 \documentclass[preprint,review,12pt]{elsarticle}

\usepackage{amssymb}
\usepackage{amsmath}
\usepackage{multirow}

\journal{Elsevier}

\begin{document}

\begin{frontmatter}


%%\tnotetext[label1]{}
%%\author{Name\corref{cor1}\fnref{label2}}
%%\ead{email address}
%%\ead[url]{home page}
%%\fntext[label2]{}
%%\cortext[cor1]{}
%%\affiliation{organization={},
%%\fntext[label3]{}

\title{An Active Fault-Tolerant Online Control Allocation Scheme for a Dual-System UAV in Transition Flight}

\author[1]{Junfeng Cai\corref{cor1}}
\ead{junfeng.cai@polimi.it}

\author[1]{Marco Lovera}
\ead{marco.lovera@polimi.it}

\cortext[cor1]{Corresponding author}

\affiliation[1]{organization={Department of Aerospace Science and Technology, Politecnico di Milano},%Department and Organization
	addressline={Via La Masa 34}, 
	city={Milan},
	postcode={20156}, 
	state={MI},
	country={Italy}}

%% Abstract
\begin{abstract}
A novel active fault-tolerant control (AFTC) scheme for a dual-system vertical takeoff and landing (VTOL) unmanned aerial vehicle (UAV) during transition flight is proposed in this paper. The AFTC scheme is composed of a baseline control law and an online control reallocation module. First, the structured $H_{\infty}$ baseline control law is able to guarantee the stability of closed-loop systems without being reconfigured under simultaneous actuator fault conditions. Second, compared to the existing mainstream method of sliding mode control that is a discontinuous control strategy, the AFTC scheme can effectively avoid control chattering problem by adopting the structured $H_{\infty}$ baseline control law. Third, an online control allocation (CA) module is implemented to carry out a unified CA for all the available actuators. When actuator faults/failures occur, the CA matrix is updated according to fault information and real-time airspeed, which is able to redistribute the virtual control signals to the remaining healthy actuators, avoiding significant performance degradation. Based on the developed AFTC scheme, symmetric and non-symmetric actuator fault scenarios are simulated on a nonlinear six-degree-of-freedom simulator, where the cases of merely structured $H_{\infty}$ control and structured $H_{\infty}$ based AFTC are compared and analyzed. The results show that the proposed structured $H_{\infty}$ based AFTC system is capable of handling more complicated fault scenarios and model uncertainties with no need to reconfigure the baseline control law. The proposed AFTC scheme significantly improves the safety and reliability of the transition flight of dual-system VTOL UAVs.
\end{abstract}

%% Keywords

\end{frontmatter}




\section{Introduction}
In recent years, unmanned aerial vehicles (UAVs) have been widely applied to various fields, including farming, forest fire monitoring, logistics and rescue activities and so on. They are becoming the main focus of researchers around the world. Among them, hybrid VTOL UAVs have particular advantages compared to conventional multicopters and fixed-wing aircraft. That is, hybrid VTOL UAVs are able to take off and land without runways unlike fixed-wing aircraft and fly longer distances than multicopters. Hybrid VTOL UAVs can be divided into three different types, i.e., the tilt rotor UAV, the tail-sitter UAV and the dual-system UAV. The dual-system UAV is getting increasing attention because of its simpler mechanical structures and relatively smooth attitude variations during the transition flight, which is promising to be applied to air mobility and logistics. Since the potential applications of dual-system UAVs are mainly in urban areas, the flight safety and reliability of such UAVs are of significance. Therefore, development of the capability of dual-system UAVs to maintain safe flight under faulty conditions, i.e., fault-tolerant capability, is meaningful for improving the reliability of transition flights.  

A complete flight mission of a hybrid VTOL UAV usually consists of various flight phases, such as multicopter flight, transition flight and fixed-wing flight. The transition flight is the most complicated flight phase. The control development for the transition flight of hybrid VTOL UAVs is challenging. In this regard, there have been some existing results. In \cite{ref1}, a nonlinear robust controller was designed for the transition flight of a tail-sitter aircraft, which involves a nominal $H_{\infty}$ controller based on time-invariant terms of a nonlinear model and a nonlinear disturbance observer. In \cite{ref2}, a transition flight controller for a Canard Rotor/Wing UAV was developed using adaptive neural network dynamic inversion by simplifying the transition flight to the longitudinal plane. In \cite{ref3}, an incremental nonlinear dynamic inversion controller combined with an incremental control allocation method was proposed for the transition flight of a tilt-rotor VTOL aircraft. A linear active disturbance rejection controller integrated with control allocation was employed to control the transition flight of a quad-plane VTOL UAV, where different control allocation (CA) methods were applied to compare the power efficiency during the transition \cite{ref4}. However, the existing research work mainly focuses on the control of the transition flight, which does not involve the fault-tolerant aspect of dual-system UAV transition flight. Additionally, in \cite{ref2,ref3} the modeling of the transition flight process is simplified to the longitudinal plane, thus limiting the discussion to longitudinal motion, and lateral and directional motions are not considered. To this end, the fault-tolerant control of dual-system UAV transition flight based on a multi-channel control design remains to be further investigated.

The dual-system UAV has a hybrid configuration with vertical and horizontal rotors, fixed wings as well as aerodynamic control surfaces. During the transition flight, all the actuators are activated. The dual system UAV is a typical over-actuated system during the transition, which provides the possibility of developing the fault-tolerant capability of the control system. Fault-tolerant control (FTC) aims at achieving safe flight after fault occurrence by means of designing fault-tolerant control systems. FTC can be categorized into passive FTC and active FTC. Where passive FTC (PFTC) consists in designing fault-tolerant control systems that do not require fault information. Instead, active FTC (AFTC) is aimed at developing fault-tolerant control systems that change control structures or parameters in real time with the help of fault information under faulty conditions. 

CA-based fault-tolerant control, as one of the most important FTC methods, is widely applied to aerospace systems. First, control allocation \cite{ref5} is an effective way to manage actuator redundancy, the application of which leads to many interesting results related to over-actuated systems. In \cite{ref6,ref7,ref8,ref9}, the idea of CA-based control design was firstly proposed and various methodologies for solving CA problems were developed, providing a basis for the following research. Quadratic programming-based control allocation \cite{ref10}, Pseudo inverse-based control allocation \cite{ref11}, Dynamic control allocation \cite{ref12} already have successful applications in flight control, based on which CA-based fault-tolerant control is further developed. CA-based FTC consists in eliminating the adverse effects caused by actuator faults/failures that occur in systems by redistributing virtual control signals (outputs of control laws) to the remaining healthy actuators. The benefits of this method are obvious. It enables a modular design of a control system that can be divided into a baseline control law and a CA module. When actuator faults/failures occur, the CA module redistributes the required control signals from the baseline control law to other healthy actuators without the need to reconfigure the baseline control law. With specific focus on FTC of hybrid VTOL UAVs, there has been some existing work related to CA-based FTC. In \cite{ref12}, an adaptive sliding mode FTC scheme based on dynamic CA for a hybrid VTOL canard rotor/wing UAV in transition flight was proposed by simplifying the transition flight to the longitudinal plane. The proposed control scheme can handle simultaneous faults of the elevator and canard. With respect to the FTC of dual-system UAVs, an incremental adaptive sliding mode control method was applied to a quad-plane aircraft in the multicopter mode considering a complete rotor failure in \cite{ref13}. In \cite{ref14}, modeling and FTC of a quad-plane UAV in the fixed-wing mode were investigated, where the FTC system is composed of a dynamic inversion control law and a control allocator. The motion of the UAV was restricted to the longitudinal plane and the fault scenarios of elevator floating and jamming were discussed. In \cite{ref15}, a sliding mode FTC design of longitudinal motion based on a LPV model was presented for a hybrid octoplane UAV in the transition mode by assuming symmetric failures of the vertical rotors and elevator. In summary, CA-based FTC is an effective approach for developing FTC systems for dual-system UAVs. At the same time, the existing research also has disadvantages and limitations in several aspects. First, the mainstream approach for hybrid VTOL UAV FTC design is based on the sliding mode control method. However, the sliding mode control method, as a discontinuous control strategy, usually causes the control chattering phenomenon, which can be clearly observed from the simulation results in \cite{ref12}. In some cases, sliding mode control based FTC can not guarantee the stability of closed-loop systems in the presence of simultaneous actuator faults/failures \cite{ref16} such that real-time control parameter update is necessary, causing undesirable transients. Second, in the existing references, FTC of dual-system UAV transition flights is investigated by considering merely the longitudinal dynamics, which limits the discussion to the longitudinal motion of aircraft. Consequently, the developed FTC systems can only deal with symmetric faults/failures that only affect the longitudinal motion. But in practical situations, symmetric actuator faults/failures rarely occur. The existing FTC research results of dual-system UAVs are only applicable to limited fault scenarios.   

Structured $H_{\infty}$ as a robust control synthesis method that is able to compute the control parameters of predefined control architectures has been used in the FTC of fixed-wing aircraft. In \cite{ref17}, structured $H_{\infty}$ was applied to the longitudinal control design of a passenger aircraft, which demonstrated the robust stability and performance of the controller under model parameter uncertainties and actuator faults/failures. A FTC strategy for a high altitude long endurance aircraft was proposed based on fault detection and isolation algorithms and a switching control law, where the switching control law was designed using the multi-model-based structured $H_{\infty}$ approach \cite{ref18}. In addition, the structured $H_{\infty}$ was also applied to flight control of multicopter UAVs. In \cite{ref19}, structured $H_{\infty}$ control was utilized to design the helicopter flight control laws with rotor state feedback, which enabled the control system to tolerate the rotor state sensor failures. But actuator faults/failures are not involved in this research. In \cite{ref20}, a gain-scheduling FTC approach based on structured $H_{\infty}$ synthesis was proposed for a multicopter UAV. Simulation and experimental results demonstrated the validity and robustness of the proposed control approach in the presence of unknown parameters and rotor failures. In \cite{ref21}, an adaptive CA scheme in the framework of structured $H_{\infty}$ was presented for a hexacopter using LQR control as a baseline control law, indicating the effectiveness and the robustness of the proposed FTC system. The existing references show that the structured $H_{\infty}$ is an effective and promising FTC approach. Whereas this method has not been applied to the FTC design of dual-system UAVs. Hence, it is interesting to investigate the applicability of the structured $H_{\infty}$ control reallocation FTC design to dual-system UAV transition flight.

This paper is devoted to the development of a novel active FTC scheme for the dual-system UAV transition flight, which is the combination of structured $H_{\infty}$ control and online control allocation. Considering the features of the transition flight, the mathematical models of altitude and full attitude dynamics are established first, based on which the corresponding control-oriented models are obtained. Then the baseline control law is designed by the structured $H_{\infty}$ leading to nominal $H_{\infty}$ controllers with a P-PID structure. Accordingly, the stability of the closed-loop systems with loss of control effectiveness caused by actuator faults/failures is analyzed, showing that the closed-loop stability is guaranteed under simultaneous actuator faults/failures. Subsequently, the quadratic programming optimization-based online control allocation, by including both the vertical rotors and all the aerodynamic control surfaces in the allocation matrix, is further described. Finally, the simulation results are reported considering two different fault scenarios involving both symmetric and non-symmetric actuator faults/failures. The simulation results demonstrate the effectiveness of the proposed structured $H_{\infty}$ control reallocation FTC system under complicated fault scenarios by comparison with the case without reallocation.

The contributions of this paper compared to the existing research work are threefold: \\
\begin{enumerate}
	\item This study is the first attempt of the structured $H_{\infty}$-based AFTC for the dual-system UAV transition flight. The structured $H_{\infty}$ as the baseline control law is able to overcome the disadvantage of the current mainstream sliding mode control methodology, i.e., control chattering problem (as shown in \cite{ref12}), resulting from the discontinuous control strategy. 
	\item Sliding mode control as a baseline control law can not maintain the stability of closed-loop systems in some cases, such as simultaneous actuator faults/failures, such that the reconfiguration of baseline control laws is inevitable \cite{ref12,ref16}. In our research, the stability of the structured $H_{\infty}$-based AFTC system is ensured under simultaneous actuator faults/failures without the need of reconfiguring the baseline control law, avoiding undesirable transients caused by the reconfiguration.
	\item The AFTC system proposed in this paper is capable of handling more complicated fault scenarios and model uncertainties. First, we consider a multi-channel FTC design including the altitude and attitude control loops instead of simplifying the transition process to the longitudinal plane, which allows the FTC system to deal with non-symmetric faults/failures that will influence not only the longitudinal motion but also the lateral and directional motions. Second, the capability of the proposed AFTC scheme to deal with complicated fault scenarios is also realized by including both the vertical rotors and the aerodynamic control surfaces in the CA matrix to perform a unified CA for all the actuators available. This allows the ailerons and rudders to be activated in the case of non-symmetric faults/failures. In addition, since there are neglected terms in the control-oriented models compared to the nonlinear dynamics, the AFTC system is robust to the model uncertainties.     
\end{enumerate}

The remainder of this paper is organized as follows. Section \ref{Sec.2} introduces our research object of the dual-system UAV and corresponding equations of motion, based on which the control-oriented models are further presented. Section \ref{Sec.3} presents the structured $H_{\infty}$-based AFTC scheme, which can be divided into the baseline controller synthesis and the implementation of the online control allocation module. Section \ref{Sec.4} reports the simulation results of the structured $H_{\infty}$-based AFTC scheme in comparison with only the structured $H_{\infty}$ control under both symmetric and non-symmetric actuator faults/failures. Section \ref{Sec.5} draws conclusions and concludes the paper.

\section{Problem Formulation}\label{Sec.2}
In this section our research object is described first. Then the nonlinear dynamics equations of the dual-system UAV during the transition are presented and the corresponding control-oriented models are further derived, which will be utilized in the control synthesis in the following section.

\subsection{The dual-system VTOL UAV}
\begin{figure}
	\centering
	\includegraphics[width=0.9\linewidth]{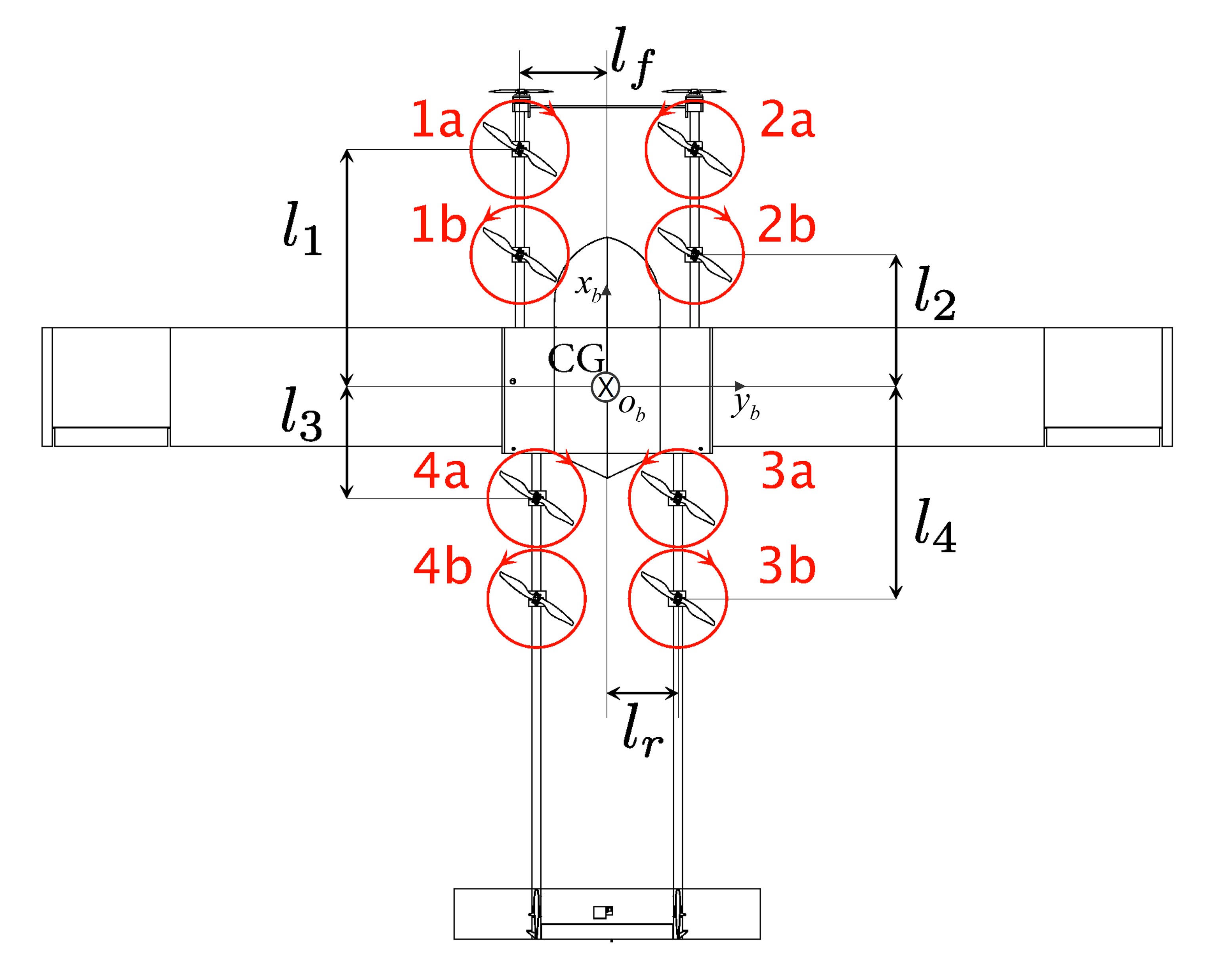}
	\caption{The top view of the dual-system UAV}
	\label{fig:figure-1}
\end{figure}

The research object of this paper is a dual-system VTOL UAV (as shown in Figure \ref{fig:figure-1}) with eight vertical rotors, two forward-mounted horizontal rotors, airframe, fixed wings, and aerodynamic control surfaces including the ailerons, elevator and rudders. The parameter set of the dual-system UAV is given in Table \ref{tab1}. 

A typical flight mission of a dual-system UAV consists of several different flight modes, i.e., the multicoper mode, fixed-wing mode and transition mode. When the VTOL is taking off or landing, it operates in the multicopter mode with only the vertical rotors remaining operational. While in the fixed-wing mode, the vertical rotors stop working, and the horizontal rotors and aerodynamic control surfaces are operating such that the UAV flies like a fixed-wing aircraft. The transition mode is an intermediate mode from the multicopter mode to the fixed-wing mode, in which all the actuators, including the rotors and the aerodynamic control surfaces, are activated and managed by the CA module.  

\begin{table}
	\begin{center}
		\caption{The parameters of the dual-system UAV}
		\label{tab1}
		\begin{tabular}{| c | c |}
			\hline
			Parameter                                                             & Value   \\
			\hline
			Mass, $m$ (kg)                                                        & 6.4     \\
			\hline
			Cruise speed, $V_{a}$ (m/s)                                           & 15      \\
			\hline
			Wing span, $b$ (m)                                                    & 2.25    \\
			\hline
			Wing area, $S_{w}$ ($\rm m^{2}$)                                      & 0.5625  \\
			\hline
			Mean aerodynamic chord, $\bar{c}$ (m)                                 & 0.25   \\
			\hline
			Inertia moment, $J$ (kg$\ \cdot\ \rm m^{2}$)                          & $ \begin{bmatrix} 0.3724    & 0     & 0.0083  \\  0     & 0.7237   & 0   \\ 0.0083   & 0     & 1.0868  \\  \end{bmatrix} $   \\
			\hline
			Radius of the horizontal                                                &\multirow{2}{*}{0.0635} \\
			propellers, $R_{h}$ (m)                                                 &  \\
			\hline
			Radius of the vertical                                                  &\multirow{2}{*}{0.0889}\\
			propellers, $R_{v}$ (m)                                                 & \\
			\hline
			Constraints on deflection                                               &\multirow{2}{*}{[-32,\ 32]} \\
			angle of the ailerons, $\delta_{a}$ (deg)                               &   \\
			\hline
			Constraints on deflection                                               &\multirow{2}{*}{[-29,\ 29]}\\
			angle of the elevator, $\delta_{e}$ (deg)                              &   \\
			\hline
			Constraints on deflection                                               & \multirow{2}{*}{[-40,\ 40]}\\
			angle of the rudders, $\delta_{r}$ (deg)                                &     \\
			\hline
		\end{tabular}
	\end{center}
\end{table}

\subsection{Nonlinear dynamics equations}
Since we assume that the baseline motion of the transition flight is steady level flight, which means that the altitude and attitude control are crucial, we mainly focus on the altitude and attitude dynamics of the dual-system UAV during the transition. The horizontal position control is not the main concern of this paper. The nonlinear equations of motion are the same as those in \cite{ref22}. 

\subsection{Control-oriented models}
Typically, the nonlinear equations of motion are used to build the nonlinear six-degree-of-freedom simulator. Whereas they are too complicated to be applied to the control synthesis. For this reason, we need simplified design models, also referred to as control-oriented models, for control development as well as for capturing the dominant features of the aircraft motion. To obtain the control-oriented models, we need to introduce some assumptions for model simplification and subsequent control design. 

\textit{Assumption 1:} The baseline motion of the dual-system UAV during the transition flight is constant altitude steady level flight.

\textit{Assumption 2:} Complete failures of at most four vertical rotors and aerodynamic control surfaces are allowed.

\textit{Assumption 3:} The actuator dynamics are sufficiently fast such that the transient processes in the actuator responses are negligible. 

According to Assumption 1, the nonlinear equations of motion are linearized at the trim conditions ($p^{*}=q^{*}=r^{*}=0$, $\phi^{*}=\psi^{*}=0$, $v^{*}=w^{*}=0$). If we further take into account the possible occurrence of actuator faults/failures, the loss of effectiveness of control forces and moments, denoted as $\Gamma =$ $[\gamma_{T}\ \gamma_{L}\ \gamma_{M}\ \gamma_{N}]$, can be included in the control-oriented models as well. Thus, the linearized control-oriented model is given as \cite{ref22}: 

\begin{equation}\label{equ.1}
	\begin{aligned}
		\dot{w} &= \frac{(1 - \gamma_{T}) F_{az}}{m} + g \\
		\dot{p} &= \frac{(1 - \gamma_{L}) M_{ax}}{J_{x}}  \\
		\dot{q} &= \frac{(1 - \gamma_{M}) M_{ay} }{J_{y}}  \\
		\dot{r} &= \frac{(1 - \gamma_{N}) M_{az}}{J_{z}}.
	\end{aligned}
\end{equation} 

Theoretically speaking, the variation range of components of $\Gamma$ is [0, 1]. Where 0 denotes the no-fault case. 1 represents a complete failure of an actuator. Since we do not discuss the case of complete failures of all the actuators, and recalling from Assumption 2, the loss of effectiveness of the control forces and moments is approximately no more than 0.5. Then it leads to the considered value ranges of the components of $\Gamma$: $\gamma_{T} \in [0,\ 0.5]$, $\gamma_{L} \in [0,\ 0.5]$, $\gamma_{M} \in [0,\ 0.5]$, $\gamma_{N} \in [0,\ 0.5]$. \\

It should be emphasized that the structured $H_{\infty}$ control synthesis in this paper is based on the nominal models. That is, the models used in the control design correspond to $\Gamma =$ $[0\ 0\ 0\ 0]$. Based on the synthesized nominal controllers, a posteriori stability analysis of the closed-loop systems will be performed by considering the possible variation range of $\gamma_{*} \in [0,\ 0.5]$.      

\section{Fault-tolerant control scheme}\label{Sec.3}
In the previous section, the nonlinear dynamics and the control-oriented models are described, which are the basis for the subsequent baseline control law synthesis. In this section, an online control allocation FTC scheme based on the structured $H_{\infty}$ control synthesis is proposed. The control architecture of the AFTC system, the baseline control law design, and the online control allocation module are described in detail.      

\subsection{Control architecture}
\begin{figure*}
	\centering
	\includegraphics[width=0.8\linewidth]{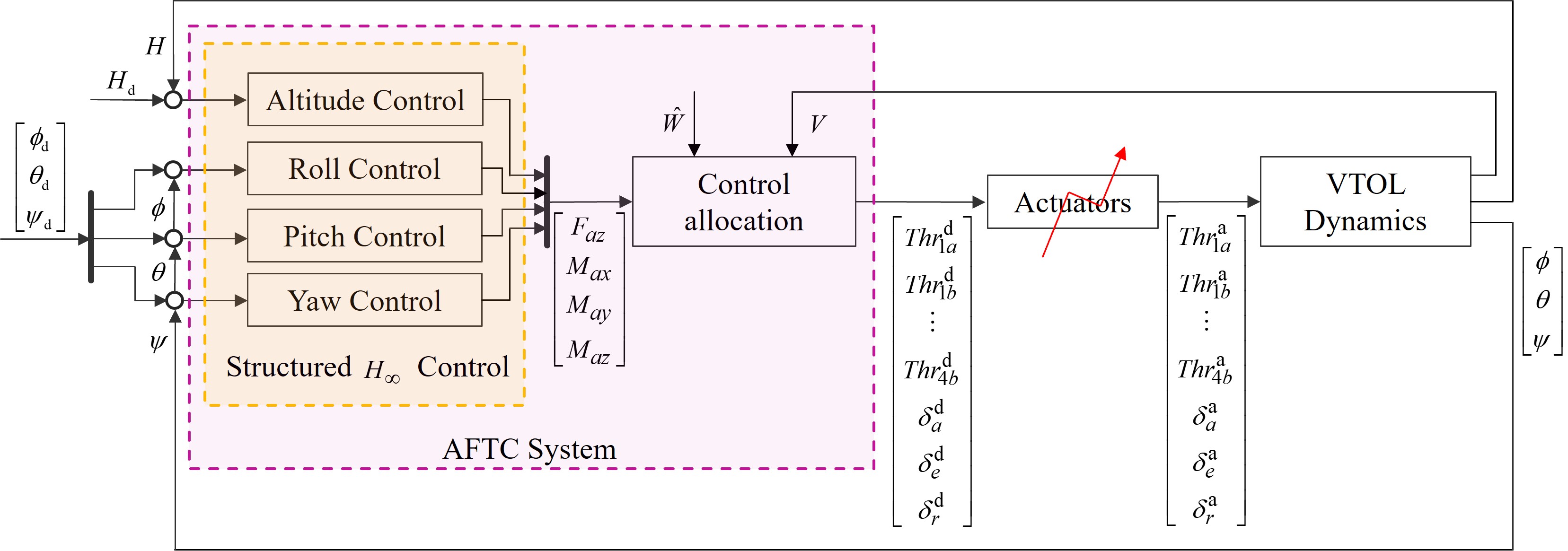}
	\caption{The architecture of the AFTC system}
	\label{fig:figure-2}
\end{figure*}

The control architecture of the structured $H_{\infty}$-based AFTC system is shown in Figure \ref{fig:figure-2}. Where the superscript 'd' denotes commanded values, the superscript 'a' denotes actual values. As can be seen, the AFTC system consists of two parts: the baseline control law and the CA module. 

The structured $H_{\infty}$ baseline control law consists of an altitude controller and three attitude controllers. The inputs to the controllers are VTOL's altitude and attitude control errors, and the outputs (known as virtual control signals) corresponding to the vertical control force and attitude control moments are sent to the online CA module. 

The CA module distributes the virtual control signals to the available actuators, and calculates the required commands for the actuators. Note that the CA matrix is airspeed-dependent. Since the airspeed of the VTOL is varying during the transition flight, the CA module computes the actuator commands by updating the airspeed values in real time, i.e., so-called online CA. When actuator faults/failures occur, an estimate of the loss of actuator effectiveness $\hat{W}$ (which will be described later) as fault information is input to the CA module. Usually, fault information is estimated by a separate module called fault detection and diagnosis (FDD). However, in this paper, the fault information is assumed to be perfectly known since FDD is beyond the scope of this paper. The CA module redistributes the virtual control signals to the remaining healthy actuators according to the fault information to eliminate the adverse effects caused by the presence of actuator faults/failures.     

\subsection{Baseline control law}
The baseline control law is developed based on the structured $H_{\infty}$ synthesis. First, the nominal controller synthesis is detailed. Then, in order to check the stability of the closed-loop systems under actuator faults/failures, a stability analysis based on a $\mu$-analysis is performed.

\subsubsection{Structured $H_{\infty}$ control synthesis}
The structured $H_{\infty}$ control has been widely used in recent years, leveraging the advantage of calculating parameters for fixed-structure controllers with specified orders. Structured $H_{\infty}$ control synthesis consists in optimizing the $H_{\infty}$ norms of the sensitivity function $S(s)$ and control sensitivity function $R(s)$ simultaneously (see \ref{fig:figure-3}). To this purpose, two different frequency dependent weighting functions $W_{s}(s)$, $W_{r}(s)$ are applied to the sensitivity function and control sensitivity function, respectively. The forms of the weighting functions are given by:

\begin{figure}
	\centering
	\includegraphics[width=0.7\linewidth]{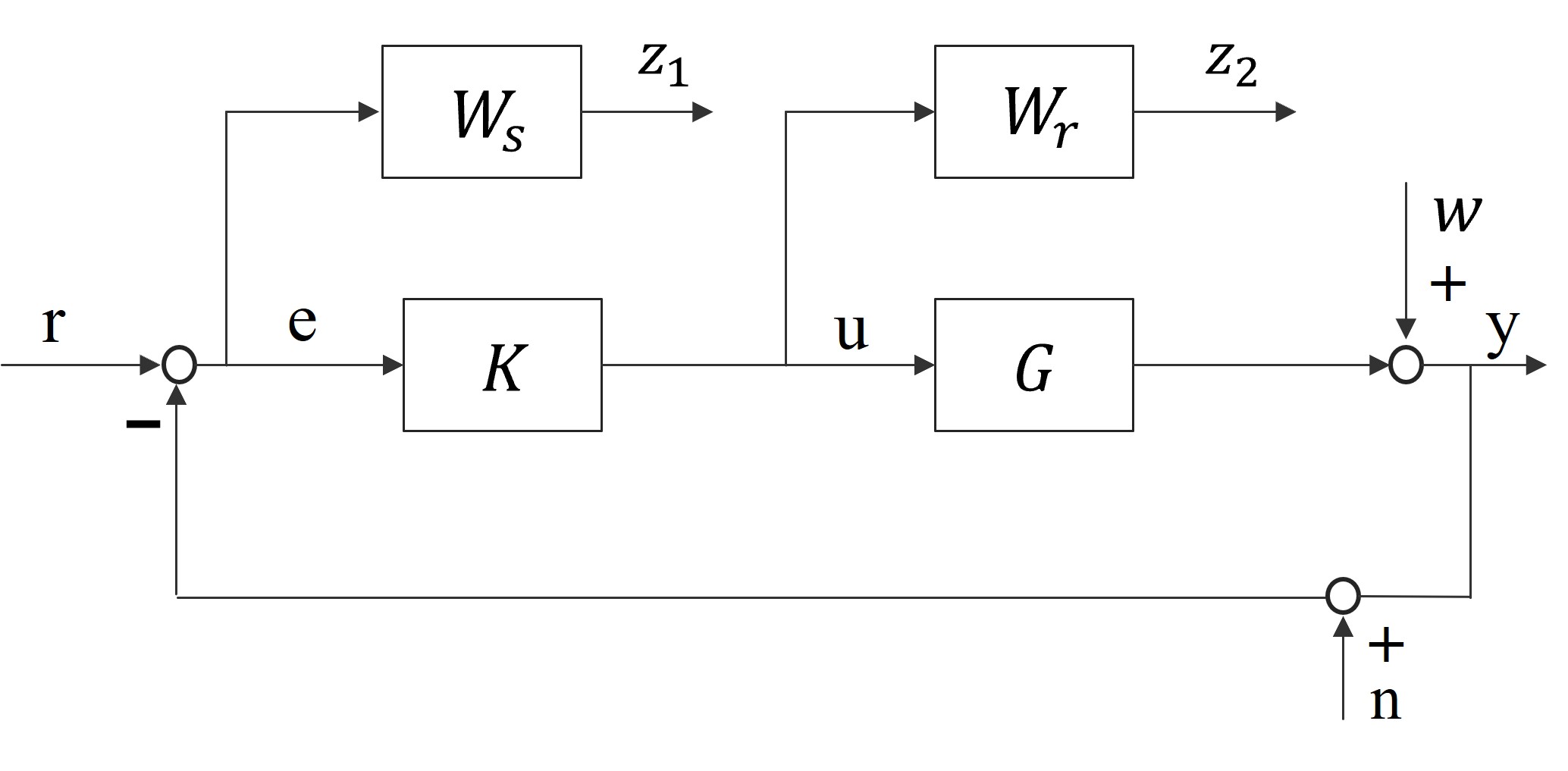}
	\caption{The general block diagram of a closed-loop system}
	\label{fig:figure-3}
\end{figure}

\begin{equation}\label{equ.4}
	W_{s}(s) = \frac{s/M + \omega_{b}}{s + A \omega_{b}}
\end{equation}         
\begin{equation}\label{equ.5}
	W_{r}(s) = \frac{r_{max}/u_{max}s + \omega_{a}10^{-3}}{s + \omega_{a}}
\end{equation}
where $M$ is the peak value of the frequency response of the reciprocal of the weighting function, which corresponds to the overshoot requirement of the time domain response of the control error, $A$ is the maximum steady-state error requirement, $\omega_{b}$ is the desirable bandwidth of the closed-loop system. $r_{max}$ is the upper limit of the reference signal, $u_{max}$ is the maximum control action within actuators' capabilities, $\omega_{a}$ is the bandwidth of an actuator. The mixed sensitivity optimization problem is to find controller $K$ such that the upper bound of the $H_{\infty}$ norms of the weighted sensitivity functions is minimized \cite{ref23}.

The control synthesis in this paper involves an altitude controller and three attitude controllers. Since the synthesis process and the weighting function selection for the attitude controllers are pretty similar to each other, only the synthesis of the altitude controller and the pitch angle controller will be described.

\begin{figure}
	\centering
	\includegraphics[width=0.9\linewidth]{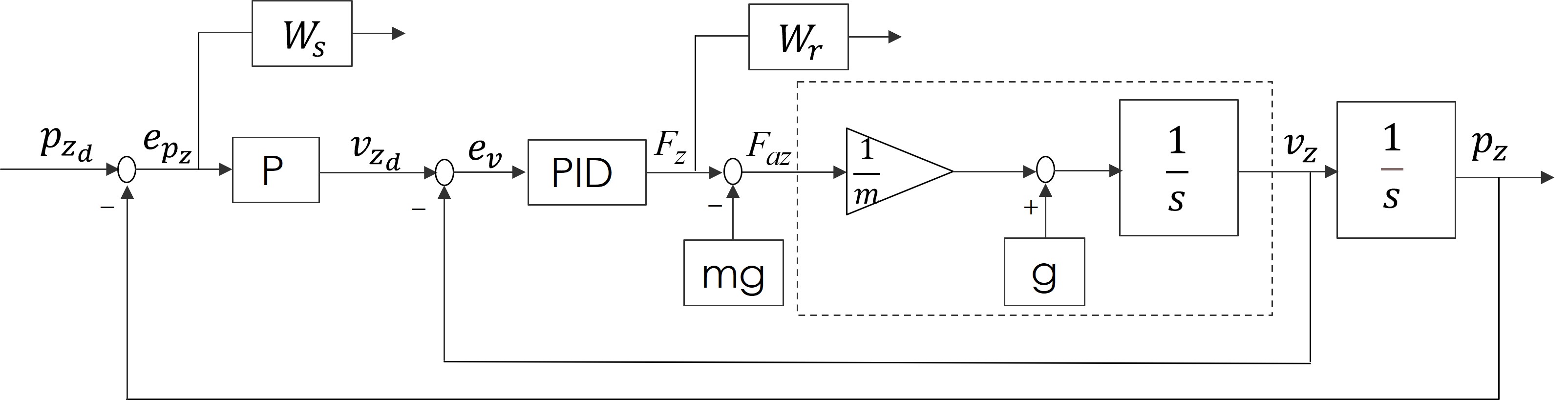}
	\caption{The structure diagram of the altitude control loop}
	\label{fig:figure-4}
\end{figure}

With respect to the altitude control, since the aerodynamic lift produced by the fixed wings is treated as an external disturbance in the control design, the altitude controller should be synthesized to be able to follow the input commands and reject the aerodynamic lift disturbance. The structure diagram is shown in Figure \ref{fig:figure-4}. It is a two-layer nested PID loop with a PID controller as the inner loop and a P controller as the outer loop. The weighting functions $W_{s}(s)$, $W_{r}(s)$ are, respectively, connected to the control error $e_{pz}$ and the resultant force in the vertical direction $F_{z}$. The parameter selection of the weighting functions is crucial to the solution of the optimization problem, which directly influences the solution results. The parameter selection is explained below:

\begin{itemize}
	\item The sensitivity weighting functions correspond to our design requirements. As shown in Equation \eqref{equ.4}, (1) $M$ is related to the overshoot requirement of time domain responses, which is set to less than 20$\%$. Accordingly, $M$ is chosen as 1.096. (2) The $A$ corresponding to the requirement for steady-state error is set to 0.001. $\omega_{b}$ is the bandwidth of the altitude control loop, and is selected as 0.8 rad/s corresponding to a settling time of 6 s. 
	\item As for the parameters in the control sensitivity weighting functions shown in Equation \eqref{equ.5}, they are determined according to their physical meaning already explained, where $\omega_{a}$ as the bandwidth of the actuators is set to 5 rad/s.
\end{itemize}

\begin{figure}
	\centering
	\includegraphics[width=0.9\linewidth]{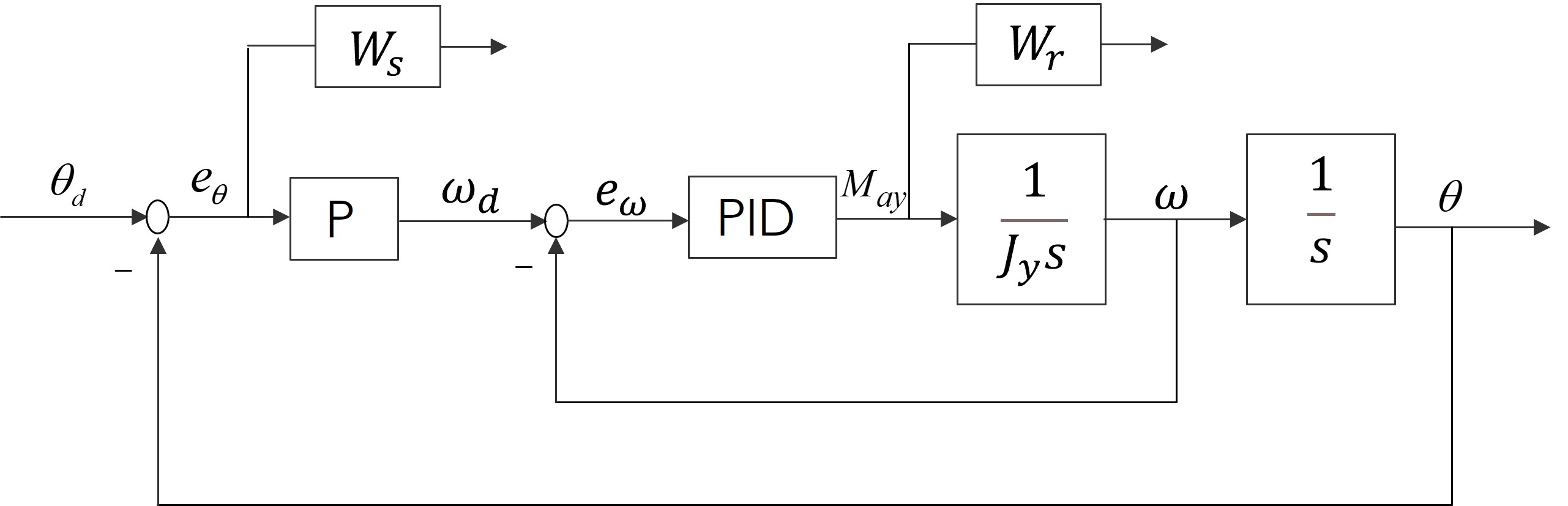}
	\caption{The structure diagram of the pitch control loop}
	\label{fig:figure-5}
\end{figure}

Regarding the pitch angle control loop, the control block diagram is shown in Figure \ref{fig:figure-5}. The weighting functions $W_{s}(s)$, $W_{r}(s)$ are applied to the pitch control error $e_{\theta}$ and the pitch control moment $M_{ay}$, respectively. In the sensitivity weighting function $W_{s}(s)$, the bandwidth of the pitch control loop $\omega_{b}$ is determined as 4.3 rad/s according to a settling time requirement of 1.2 s. The other parameters are chosen to be the same as those of the altitude control loop resulting from the same design requirements.    

Note that the design requirements above mentioned are the performance requirements for the nominal controllers. If actuator faults/failures occur, it may cause severe performance degradation of the control system. In this case, control reallocation plays an important role in compensating for the mismatches between the virtual control signals from the baseline control law and the actual control efforts that actuators can provide, which can eliminate the adverse effects to the maximum extent and thus avoid significant performance degradation, which will be shown in subsequent simulation results.

\subsubsection{A posteriori stability analysis with actuator faults/failures}
In the preceding subsection, a nominal control system using the structured $H_{\infty}$ synthesis is obtained without considering possible actuator faults/failures. But the control system is likely to lose stability when actuator faults/failures occur. Nevertheless, the stability of the synthesized controllers under actuator faults/failures is unclear to us. To this end, a posteriori stability analysis is conducted to examine the stability of the closed-loop systems after the presence of actuator faults/failures. In this paper, the structured singular value (also known as the $\mu$-analysis), as a powerful tool to analyze the stability of closed-loop systems with given control parameters under model uncertainties, is employed. 

\begin{figure}
	\centering
	\includegraphics[width=0.7\linewidth]{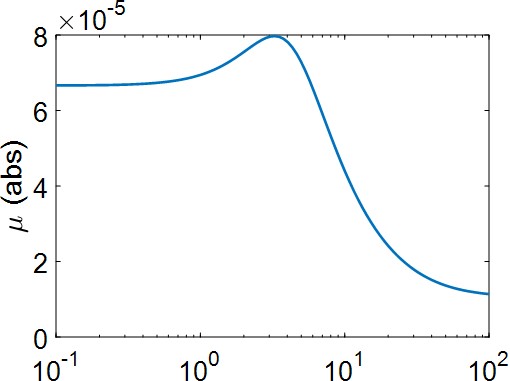}
	\caption{The $\mu$-plot of the pitch angle control loop.}
	\label{fig:figure-6}
\end{figure}

\begin{figure}
	\centering
	\includegraphics[width=0.7\linewidth]{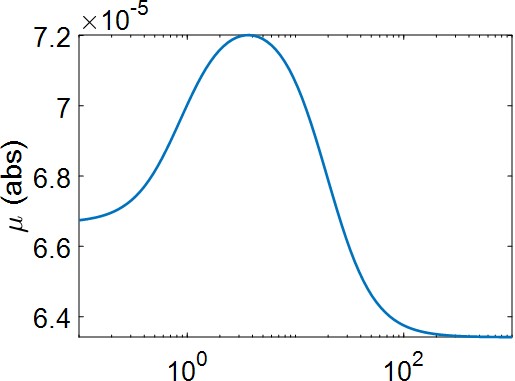}
	\caption{The $\mu$-plot of the altitude control loop.}
	\label{fig:figure-7}
\end{figure}

Since the $\mu$-plots of the attitude control loops are pretty similar to each other, to avoid repetition, only the $\mu$-plots of the altitude and pitch angle control loops are presented here (as shown in Figures \ref{fig:figure-6} and \ref{fig:figure-7}). As we can see, the maximum values of both the $\mu$-plots for altitude and pitch angle control loops are at the 10$^{-5}$ level. According to the stability condition based on the $\mu$-analysis \cite{ref23}, if the $\mu$ value of the closed-loop system is smaller than 1, then the closed-loop system is robustly stable. The $\mu$ values are far less than 1 (as shown in the $\mu$-plots), which indicates that the closed-loop systems are stable and have big stability margin under the simultaneous faults/failures of at most four vertical rotors and aerodynamic control surfaces. 

\subsection{Online control allocation}
CA is an efficient way to manage actuator redundancy in systems, allowing us to take into account actuator constraints simultaneously. As a module independent of a baseline controller, the CA module is responsible for distributing the virtual control signals output from the baseline controller to all available actuators. Especially in case of actuator faults/failures, the CA module also provides the possibility of reallocation to take full advantage of actuator redundancy, thus leading to the FTC capabilities of control systems. In our work, given the hybrid configuration of the dual-system UAV including two different types of actuators, we include all the actuators in the CA matrix to achieve a unified CA simultaneously. In addition, resorting to online update of the airspeed, the CA matrix is able to deal with the allocation problem with varying airspeed during the transition flight.   

\subsubsection{Actuator models with faults/failures}
To formulate the CA problem, firstly we should make it clear that the relationships between the virtual control forces and moments and the actuator commands, which are related to the actuator models. For the rotors, the forces and moments have approximately linear relationships with the commanded throttles, as given by:
\begin{equation}\label{equ.10}
	\begin{aligned}
		F_{i*} & = k_{T} \cdot Thr_{i*} \\
		M_{i*} & = k_{M} \cdot Thr_{i*}
	\end{aligned}
\end{equation} 
where $i = 1, 2, 3, 4$, the subscript $*$ denotes $a$ or $b$. $Thr_{i*}$ is the commanded throttle percentage, $k_{T}, k_{M}$ are coefficients obtained by calculating the identification results of the rotor dynamics. In this research, $k_{T}$ = 0.164 N/\%, $k_{M}$ = 1.89 $\times$ 10$^{-3}$ (N$\cdot$ m)/\%.

As for the aerodynamic control surfaces, the control moments produced by the control surface deflections, denoted as functions of the airspeed, can be expressed as:
\begin{equation}\label{equ.11}
	\begin{aligned}
		M_{a}(V) & = q(V)SbC_{L_{\delta_{a}}}(V) \delta_{a} \\
		M_{e}(V) & = q(V)S\bar{c}C_{M_{\delta_{e}}}(V) \delta_{e}\\
		M_{r}(V) & = q(V)SbC_{N_{\delta_{r}}}(V) \delta_{r}
	\end{aligned}	
\end{equation}  
where $C_{L_{\delta_{a}}}$ is the aerodynamic derivative of the pitch control moment with respect to the elevator deflection. $C_{M_{\delta_{e}}}$ denotes the aerodynamic derivative of the roll control moment with respect to the aileron deflection. $C_{N_{\delta_{r}}}$ represents the aerodynamic derivative of the yaw control moment with respect to the rudder deflection. $V$ is the airspeed of the dual-system UAV during the transition.

Considering possible actuator faults/failures, we introduce an actuator effectiveness matrix $W =$diag$([w_{1}\ w_{2}\ ...\ w_{11}])$ to represent the actuator effectiveness level. Where $w_{*} \in [0\ 1]$, 1 means no actuator fault whereas 0 represents a complete failure, $w_{*} \in (0\ 1)$ corresponds to partial loss of an actuator. With reference to \cite{ref22}, the forces and moments generated by the actuators integrated with the actuator effectiveness matrix can be rewritten in matrix form:
\begin{equation}\label{equ.12}
	\begin{aligned}	
		[F_{a_{z}}\ M_{a_{x}}\ M_{a_{y}}\ M_{a_{z}}]^{T} = B W [Thr_{1a}\ Thr_{1b}\ Thr_{2a}\ Thr_{2b}&\\
		Thr_{3a}\ Thr_{3b}\ Thr_{4a}\ Thr_{4b}\ \delta_{a}\ \delta_{e}\ \delta_{r}]^{T}&
	\end{aligned}	
\end{equation}
subject to constraints: $Thr_{i*} \in [0,\ 100]$, $\delta_{a} \in [-0.55,\ 0.55]$, $\delta_{e} \in [-0.5,\ 0.5]$, $\delta_{r} \in [-0.69,\ 0.69]$. Where the CA matrix $B$ is given by Equation \eqref{equ.13}, in which $M_{a}^{\delta_{a}}(V)=q(V)SbC_{L_{\delta_{a}}}(V)$, $M_{e}^{\delta_{e}}(V)=q(V)S\bar{c}C_{M_{\delta_{e}}}(V)$, $M_{r}^{\delta_{r}}(V)=q(V)SbC_{N_{\delta_{r}}}(V)$.  

\begin{figure*}
	\begin{equation}\label{equ.13}
		B = 
		\left[ 
		\begin{matrix} 
			-k_{T}       & -k_{T}       & -k_{T}        & -k_{T}        & -k_{T}        & -k_{T}          \\
			k_{T}l_{f}   & k_{T}l_{f}   & -k_{T}l_{f}   & -k_{T}l_{f}   & -k_{T}l_{f}   & -k_{T}l_{f}     \\ 
			k_{T}l_{1}   & k_{T}l_{2}   & k_{T}l_{1}    & k_{T}l_{2}    & -k_{T}l_{3}   & -k_{T}l_{4}     \\   
			-k_{M}       &k_{M}         &k_{M}          &-k_{M}         & k_{M}         &-k_{M}           \\        
		\end{matrix} 
		\right.
		\\
		\left.
		\begin{matrix} 
			-k_{T}        & -k_{T}       & 0                     & 0                           & 0  \\
			k_{T}l_{f}    &  k_{T}l_{f}  & M_{a}^{\delta_{a}}(V)              & 0                           & 0  \\ 
			-k_{T}l_{3}   & -k_{T}l_{4}  & 0                     & M_{e}^{\delta_{e}}(V)                    & 0  \\   
			-k_{M}        & k_{M}        & 0                     & 0                           & M_{r}^{\delta_{r}}(V) 
		\end{matrix} 
		\right]
	\end{equation}
\end{figure*}

It is noticed that $B$ contains time-varying terms related to the airspeed $V$ during the transition. That is, the CA matrix needs to be updated at each time instant before being exploited in the calculation of allocation since the airspeed of the UAV increases steadily during the transition flight.

\subsubsection{Online solution}
The formulation of the online CA problem is given by Equation \eqref{equ.14}. Thus, the essence of the CA problem consists in solving this equation in real time. However, the solution of the equation that satisfies the actuator constraints does not always exist. In this context, weighted least squares CA based on control error minimization was proposed \cite{ref24}:      
\begin{equation}\label{equ.14}
	min(\|W_{1}(u-u_{d})\|^{2} + \gamma \|W_{2}(BWu-v)\|^{2}),\ \ u_{min} \leq u \leq u_{max} 
\end{equation}
where $v$ denotes the control signals required by the controller. $u$ is the actuator command. Here $v=[F_{a_{z}}\ M_{a_{x}}\ M_{a_{y}}\ M_{a_{z}}]^{T}$, $u=[Thr_{1a}\ Thr_{1b}\ Thr_{2a}\ Thr_{2b}\ \\ Thr_{3a}\ Thr_{3b}\ Thr_{4a}\ Thr_{4b}\ \delta_{a}\ \delta_{e}\ \delta_{r}]^{T}$. $u_{d}$ is the desired control effort. $W_{1}, W_{2}$ are the weighting matrices. $\gamma\gg 1$ such that the second term in Equation \eqref{equ.14} is the dominant term to be optimized. This constrained optimization problem is solved using the active set method \cite{ref24} that is able to obtain the optimal solution after a finite number of iterations. Note that before each iteration, the CA matrix must be updated according to the current airspeed of the UAV. 

\section{Simulation results}\label{Sec.4}
Based on the developed AFTC scheme, two different fault scenarios are simulated on the six-degree-of-freedom nonlinear simulator. One fault scenario is to consider the symmetric actuator faults/failures affecting only longitudinal motion, i.e., the complete failures of the vertical rotors at symmetric positions. The other is the case of non-symmetric actuator faults/failures corresponding to the complete failures of the vertical rotors at non-symmetric positions, which influences not only the longitudinal motion but also the lateral and directional motions. By simulating these two different fault scenarios, it is demonstrated that the proposed AFTC system is capable of handling complicated fault scenarios. 

Moreover, since simultaneous actuator faults/failures may cause severe performance degradation of the system relying only on the structured $H_{\infty}$ control law, a comparison is made between the simulation results of the structured $H_{\infty}$ control and the structured $H_{\infty}$-based online CA scheme, which verifies that the proposed AFTC system through control reallocation can effectively avoid severe performance degradation.      

\subsection{Symmetric faults/failures}
In this section, a symmetric fault scenario, i.e., the complete failures of rotors 1b and 2b, 50$\%$ loss of the elevator, is considered. At the beginning of the simulation, the VTOL is hovering in the multicopter mode. It starts the transition flight since 20 s. The actuator faults/failures are injected into the system at 22 s. Then the UAV continues the transition flight under this faulty condition. After the completion of the transition flight, the UAV enters the fixed-wing mode. 

\begin{figure}
	\centering
	\includegraphics[width=0.7\linewidth]{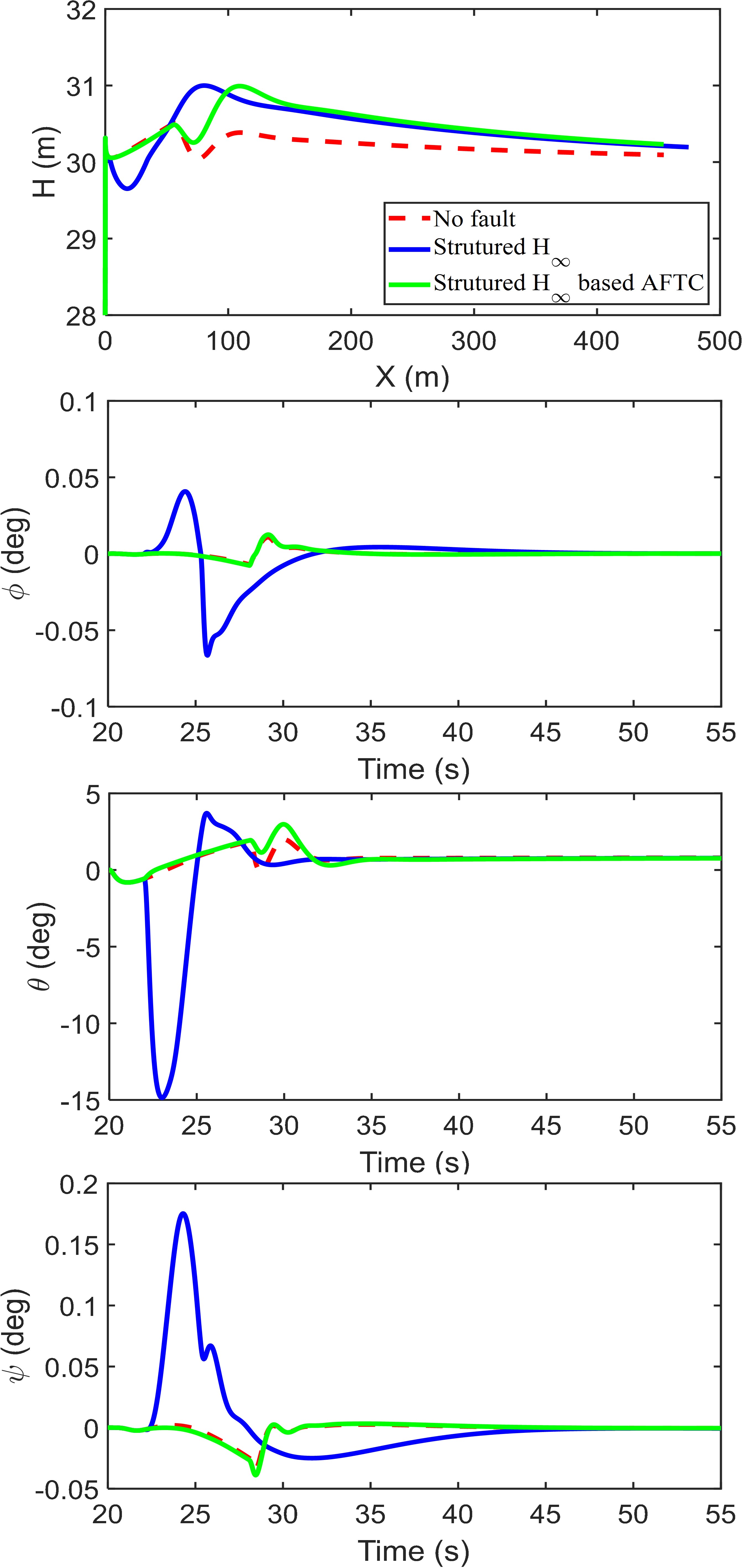}
	\caption{Trajectories in longitudinal plane and time histories of attitude angles under symmetric faults/failures. $H,\ X$ denote altitude and horizontal position. $\phi,\ \theta,\ \psi$ are roll, pitch and yaw angles, respectively.}
	\label{fig:figure-8}
\end{figure}

The trajectories in the longitudinal plane and the time histories of the attitude angles are shown in Figure \ref{fig:figure-8}. The red dashed line represents the structured $H_{\infty}$ control without actuator faults. The blue line denotes the structured $H_{\infty}$ control under actuator faults/failures, but without control reallocation. Instead, the green line corresponds to the structured $H_{\infty}$ control under actuator faults/failures with control reallocation. As can be seen from the figure of longitudinal trajectories, the maximum altitude variation is 1 m compared to the set point of 30 m.   Where during the transition flight, the longitudinal trajectory corresponding to the structured $H_{\infty}$-based AFTC coincides exactly with that of the no-fault case. The altitude using only the structured $H_{\infty}$ control decreases by 0.5 m after fault occurrence. After the completion of the transition, the altitude variations using the structured $H_{\infty}$-based AFTC are larger than those of the no-fault case. That is because during the fixed-wing flight there exists 50$\%$ loss of the elevator that is the only pitch control actuator without redundancy, thus resulting in an increase in the instantaneous vertical velocity component. 

With respect to the attitude variations, it can be seen from the figure of the pitch angle (shown in Figure \ref{fig:figure-8}) that the maximum pitch angle variation is -15 degrees after fault occurrence without control reallocation. In contrast, the figure of the pitch angle with the structured $H_{\infty}$-based AFTC is slightly different from that of the no-fault case, which effectively avoids the performance degradation through control reallocation. Regarding the roll and yaw angle variations, it is noticed that even if the assumed faults/failures are symmetric, without control reallocation the lateral and directional motions are slightly influenced as well. This is due to the fact that the coupling between the longitudinal and lateral motions is taken into account in the six-degree-of-freedom simulator, and without control reallocation there exists the errors between the required virtual control signals and the actual control efforts provided by the actuators. For the case of the structured $H_{\infty}$-based AFTC, the angle variations completely coincide with those of the no-fault case. The control reallocation perfectly compensates for the errors of the virtual control signals, which eliminates the coupling effects of the nonlinear simulator under the symmetric actuator faults/failures. 

\begin{figure}
	\centering
	\includegraphics[width=0.7\linewidth]{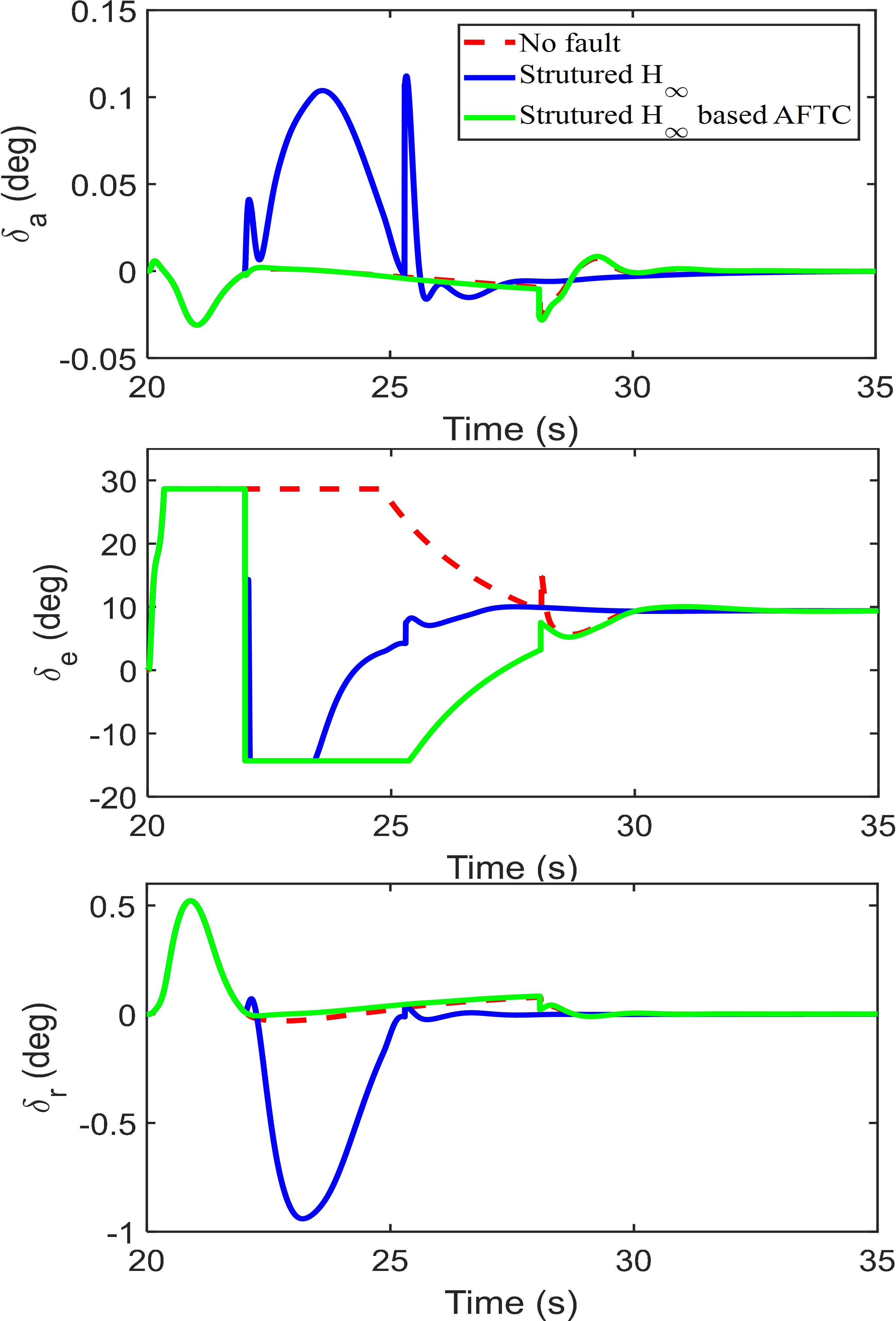}
	\caption{The variations of deflection angle of aerodynamic control surfaces. Where $\delta_{a}$, $\delta_{e}$, $\delta_{r}$ are respectively deflection angles of ailerons, elevator and rudders.}
	\label{fig:figure-9}
\end{figure}

The time histories of the aerodynamic control surface deflections are shown in Figure \ref{fig:figure-9}. For the case of the structured $H_{\infty}$ control without control reallocation, the ailerons and rudders are slightly deflected to counteract the coupling effects between different channels. The aileron and rudder deflections with the structured $H_{\infty}$-based AFTC are almost the same as those with the no-fault case, for the same reason as the roll and yaw angle variations. 

\begin{figure}
	\centering
	\includegraphics[width=1\linewidth]{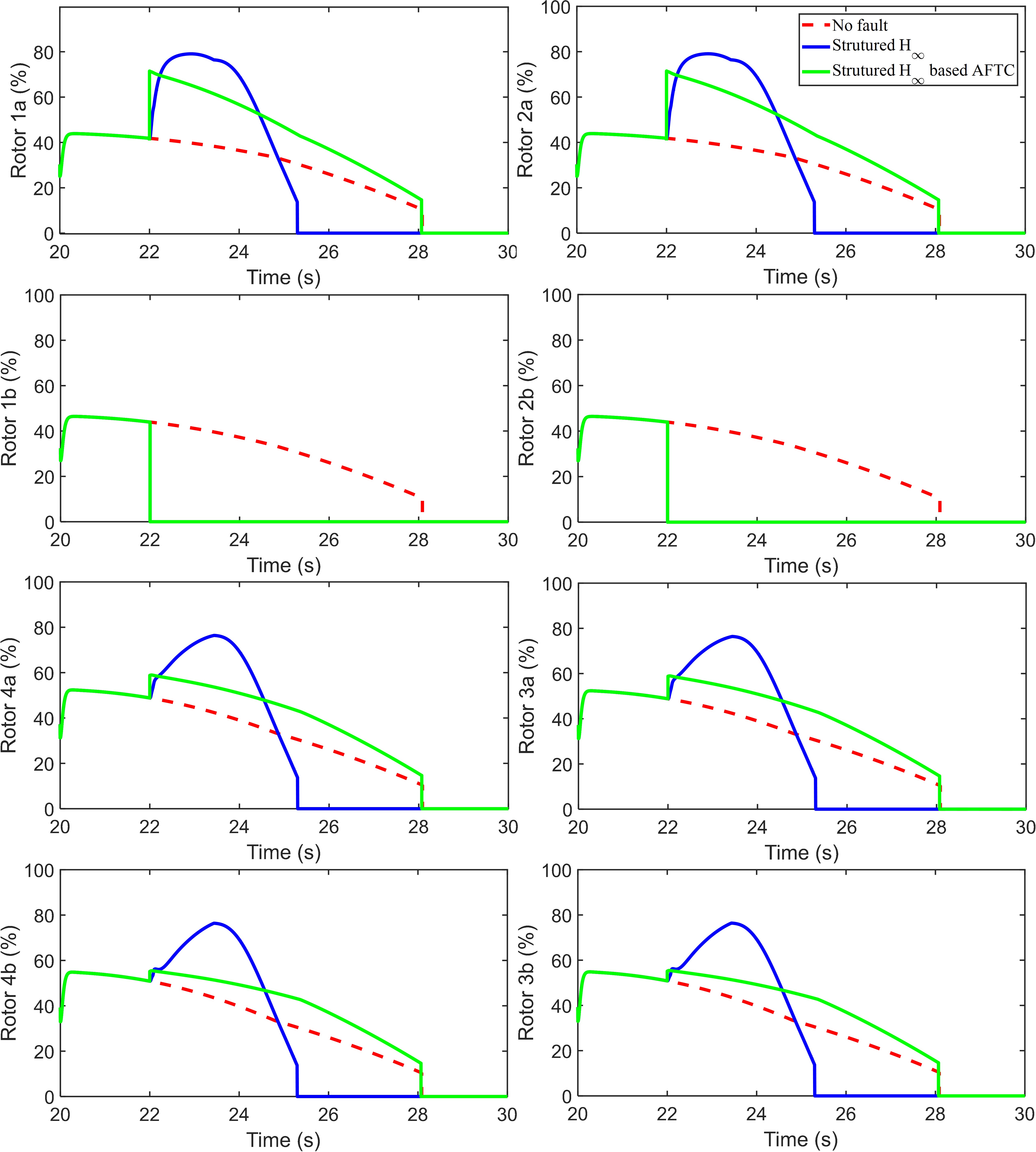}
	\caption{The variations of throttle percentage of vertical rotors.}
	\label{fig:figure-10}
\end{figure}

The time histories of the throttle percentages of the vertical rotors are shown in Figure \ref{fig:figure-10}. Note that the throttle percentages of the vertical rotors that are at the symmetric positions are identical, due to the symmetric fault assumption. For the case of the structured $H_{\infty}$-based AFTC, the throttle percentage variations are very similar to those of the no-fault case. The difference is that the throttle percentages are increased to compensate for the loss of the vertical force due to the complete failures of rotors 1b and 2b. For the case of the structured $H_{\infty}$ control only, the increases in throttle percentages are even greater, resulting in earlier completion of the transition flight. 

Thus, it is known to us that the structured $H_{\infty}$-based AFTC scheme can effectively avoid the performance degradation of the longitudinal motion, and eliminate the coupling effects between different channels by compensating for the difference between the required virtual control signals and the actual control efforts that the actuators can provide through control reallocation.     

\subsection{Non-symmetric faults/failures}
We further consider the scenario of non-symmetric actuator faults/failures. Since the non-symmetric faults/failures will cause non-symmetric control moments, to illustrate the capability of the AFTC system to cope with more complicated fault scenarios, a non-symmetric fault scenario is assumed. That is, the complete failures of rotors 1b and 3b, 50$\%$ loss of the elevator. The simulation setup is exactly the same as in the preceding subsection with merely the fault scenario changed. 

\begin{figure}
	\centering
	\includegraphics[width=0.7\linewidth]{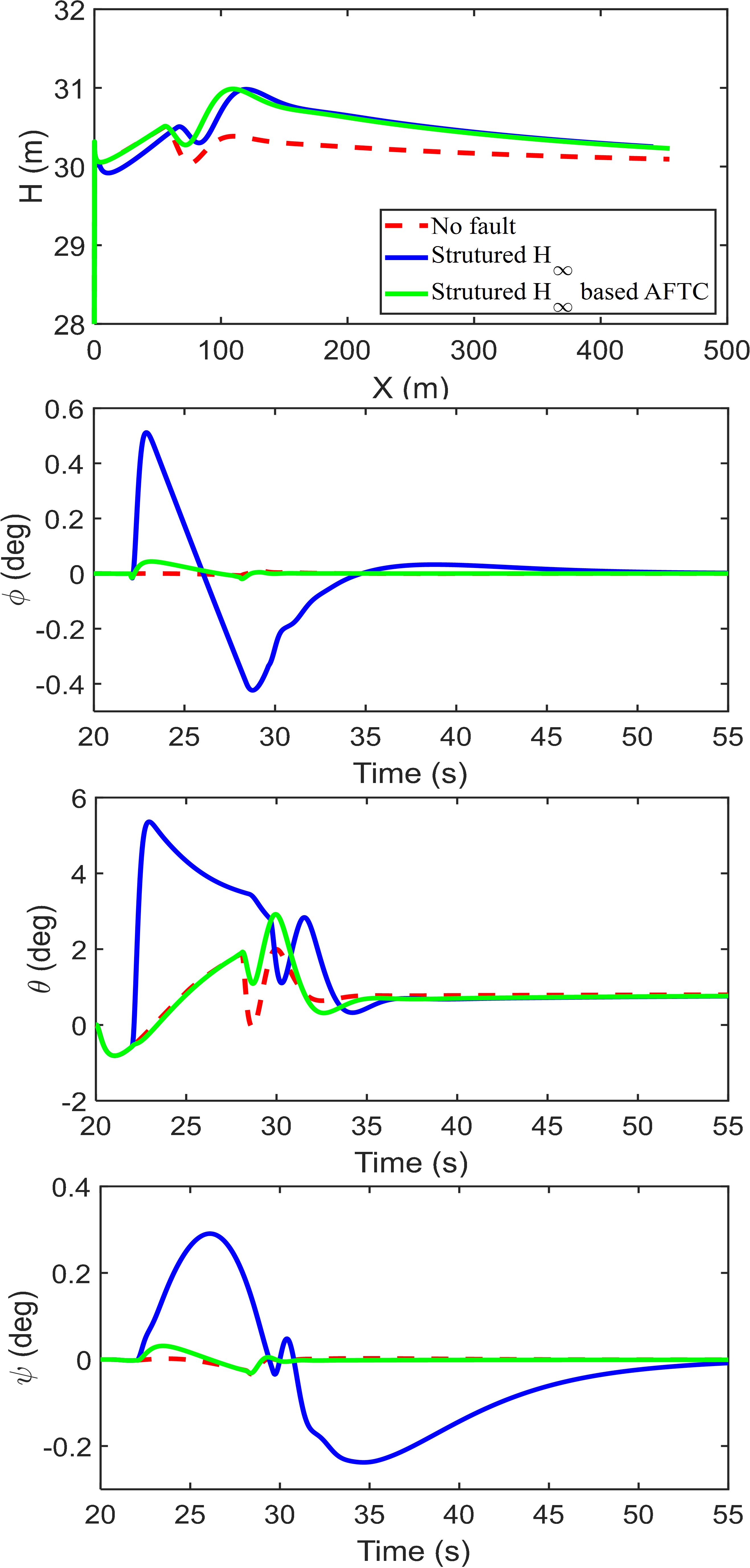}
	\caption{Trajectory in longitudinal plane and time histories of attitude angles under symmetric faults/failures. $H,\ X$ denote altitude and horizontal position. $\phi,\ \theta,\ \psi$ are respectively roll, pitch and yaw angles.}
	\label{fig:figure-11}
\end{figure}

The longitudinal trajectories and the attitude variations are shown in Figure \ref{fig:figure-11}. The longitudinal trajectories of three different cases are very similar to those of the symmetric fault scenario. 

For the pitch angle variation, the figure of the structured $H_{\infty}$-based AFTC almost coincides with that of the no-fault case. The maximum pitch angle variation corresponding to only structured $H_{\infty}$ control is 5.4 degrees. As for the roll and yaw angles, the case of structured $H_{\infty}$ control produces much larger angle variations resulting from the non-symmetric control moments. On the contrary, the structured $H_{\infty}$-based AFTC scheme can deal with the non-symmetric faults/failures with slightly increased roll and yaw angles, which obviously leads to better FTC performance than the case without control reallocation. 

\begin{figure}
	\centering
	\includegraphics[width=0.7\linewidth]{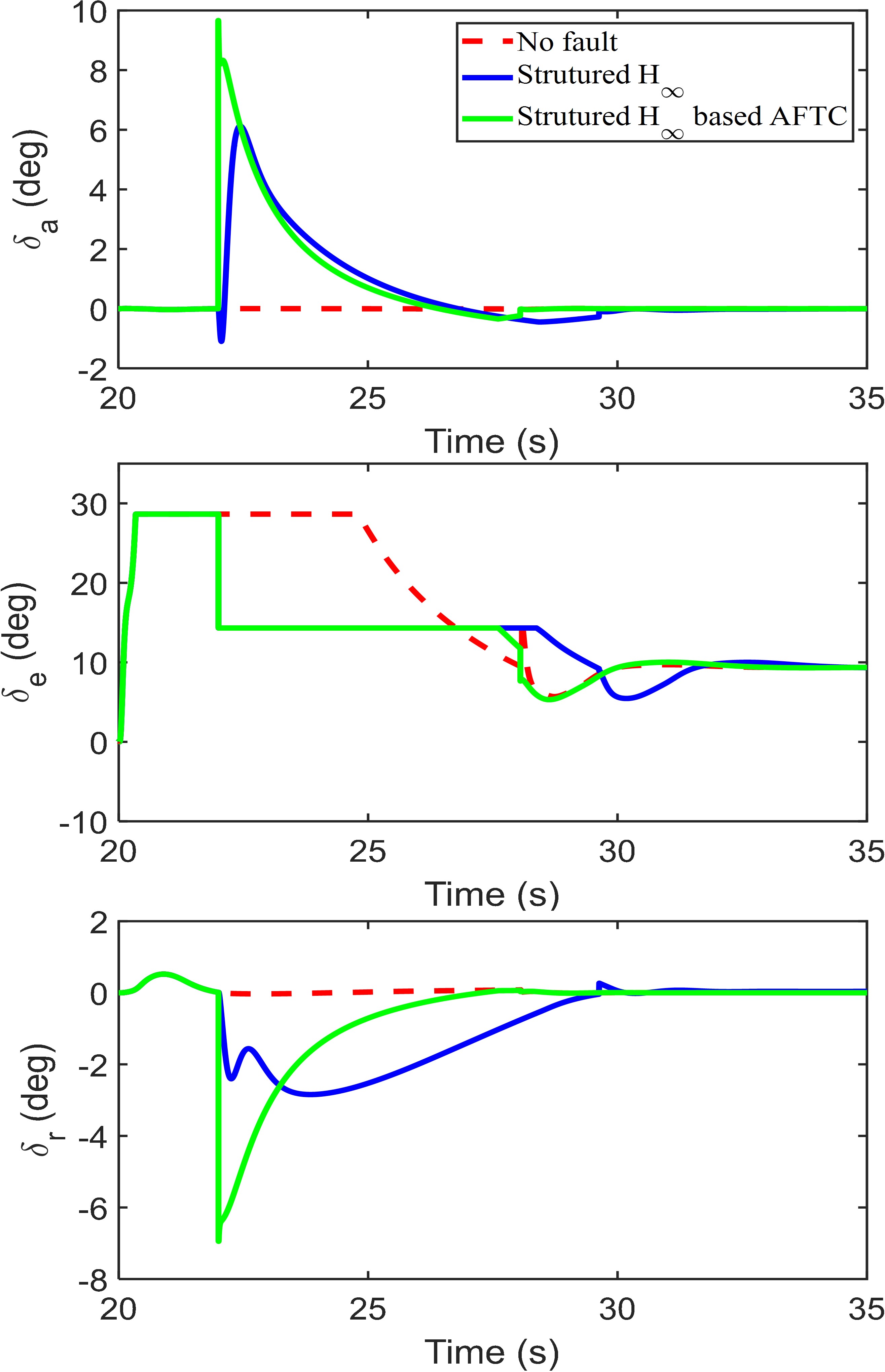}
	\caption{The variations of deflection angle of aerodynamic control surfaces. Where $\delta_{a}$, $\delta_{e}$, $\delta_{r}$ are respectively deflection angles of ailerons, elevator and rudders.}
	\label{fig:figure-12}
\end{figure}

The time histories of the aerodynamic control surface deflections are shown in Figure \ref{fig:figure-12}. It is noticed that in both the structured $H_{\infty}$ and structured $H_{\infty}$-based AFTC cases, the ailerons and rudders are activated to counteract the unbalanced roll and yaw moments due to the non-symmetric actuator faults/failures. Unlike the structured $H_{\infty}$ control, in the AFTC case, the CA module redistributes the virtual control signals to the remaining healthy actuators based on the fault information, resulting in larger aileron and rudder deflections, which takes full advantage of the redundancy of the ailerons and rudders. Thus it leads to better FTC performance. 

\begin{figure}
	\centering
	\includegraphics[width=1\linewidth]{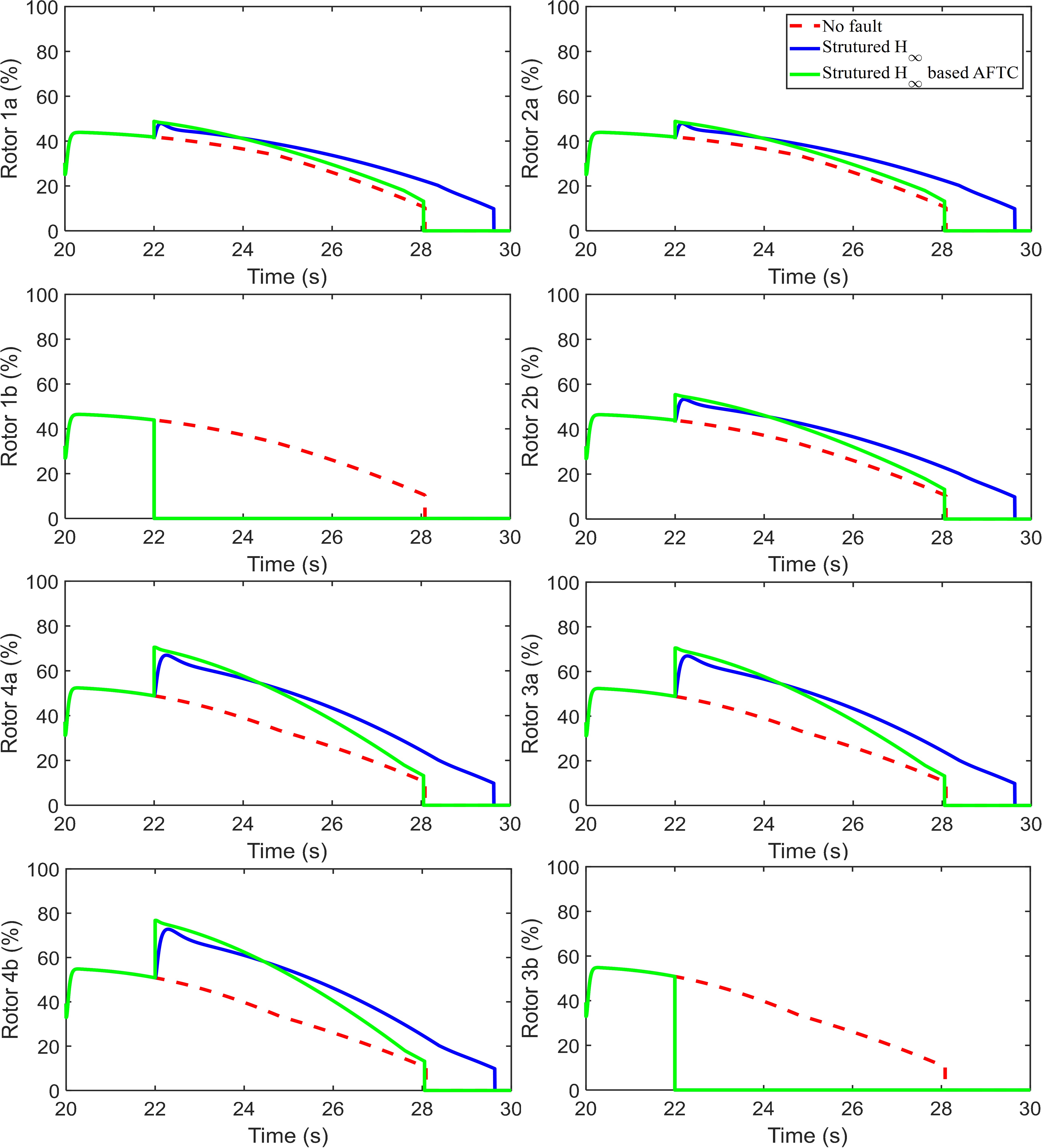}
	\caption{The variations of throttle percentage of vertical rotors.}
	\label{fig:figure-13}
\end{figure}

In Figure \ref{fig:figure-13}, the figures of the throttle percentages in the structured $H_{\infty}$-based AFTC case are pretty similar to those of the symmetric faults/failures. The throttle percentage values of the vertical rotors under both fault scenarios are almost at the same level. The structured $H_{\infty}$-based AFTC leads to the same transition flight duration as the no-fault case. Whereas the transition flight duration is longer under the structured $H_{\infty}$ control. This is because, in the structured $H_{\infty}$ case, the pitch angle is much larger than those in the structured $H_{\infty}$-based AFTC and no-fault cases, resulting in greater aerodynamic drag during transition flight. Since the critical airspeeds for the completion of the transition flight and the propulsion forces produced by the horizontal rotors are the same for all the three cases, the transition flight duration of the structured $H_{\infty}$ control is longer than those of the other two cases.                         

\section{Conclusion}\label{Sec.5}     
In this paper, a structured $H_{\infty}$-based online control reallocation FTC scheme is proposed. The developed AFTC system is composed of the structured $H_{\infty}$ baseline control law and the CA module. First, based on the control-oriented models, the nominal structured $H_{\infty}$ control synthesis including altitude and attitude control is performed without considering possible actuator faults/failures. Furthermore, the stability of the baseline controllers under the complete failures of at most four vertical rotors and aerodynamic control surfaces is evaluated using the $\mu$-analysis. Then an online CA module is implemented, which can work under both no-fault and faulty conditions.  Additionally, in order to illustrate the effectiveness of the proposed AFTC scheme, two different fault scenarios are simulated on the nonlinear six-degree-of-freedom simulator. The conclusions are as follows:

\noindent 1. The stability analysis results indicate that the structured $H_{\infty}$ controllers are able to maintain the stability of the closed-loop systems without reconfiguration under the given faulty conditions. 

\noindent 2. The structured $H_{\infty}$ control as the baseline control law has the benefit of avoiding the use of a discontinuous control strategy compared to the mainstream sliding mode control method, thus avoiding the control chattering phenomenon. 

\noindent 3. The online CA module is capable to compensate for the errors between the required virtual control signals and the actual control efforts that the actuators can provide by redistributing the virtual control signals to the remaining healthy actuators. Combined with the structured $H_{\infty}$ baseline controllers, the AFTC system is able to effectively avoid significant performance degradation compared to the variant without control reallocation.

\noindent 4. Under both symmetric and non-symmetric fault scenarios, the AFTC system exhibits much better FTC performance than the structured $H_{\infty}$ control without control reallocation, with no significant performance degradation compared to the no-fault case. The simulation results show the capability of the developed AFTC system to tolerate both symmetric and non-symmetric actuator faults/failures and model uncertainties, which further demonstrates the validity of the AFTC system in handling more complicated fault scenarios compared to the existing references.

\section*{Acknowledgments}
The first author acknowledges the financial support from China Scholarship Council (CSC, No.202006020041).

\end{document}